# Effect of high pressure annealing on the normal state transport of LaO$_{0.5}$F$_{0.5}$BiS$_2$

I. Pallecchi, G. Lamura, M. Putti
*CNR-SPIN and Università di Genova, via Dodecaneso 33, I-16146 Genova, Italy*

J. Kajitani, Y. Mizuguchi, O. Miura
*Department of Electrical and Electronic Engineering, Tokyo Metropolitan University, Hachioji, Tokyo 192-0397, Japan*

S. Demura
*Tokyo University of Science, 1-3 Kagurazaka, Shinjuku-ku, Tokyo 162-8601, Japan*

K. Deguchi, Y. Takano
*National Institute for Materials Science, Tsukuba, Ibaraki 305-0047, Japan*

## Abstract

We study normal state electrical, thermoelectrical and thermal transport in polycrystalline BiS$_2$-based compounds, which become superconducting by F doping on the O site. In particular we explore undoped LaOBiS$_2$ and doped LaO$_{0.5}$F$_{0.5}$BiS$_2$ samples, prepared either with or without high pressure annealing, in order to evidence the roles of doping and preparation conditions. The high pressure annealed sample exhibits room temperature values of resistivity ρ around 5 mΩcm, Seebeck coefficient S around -20 μV/K and thermal conductivity κ around 1.5 W/Km, while the Hall resistance $R_H$ is negative at all temperatures and its value is -10$^{-8}$ m$^3$/C at low temperature. The sample prepared at ambient pressure exhibits $R_H$ positive in sign and five times larger in magnitude, and S negative in sign and slightly smaller in magnitude. These results reveal a complex multiband evolution brought about by high pressure annealing. In particular, the sign inversion and magnitude suppression of $R_H$, indicating increased electron-type carrier density in the high pressure sample, may be closely related to previous findings about change in lattice parameters and enhancement of superconducting $T_c$ by high pressure annealing. As for the undoped sample, it exhibits the 10 times larger resistivity, 10 times larger |S| and 10 times larger |$R_H$| than its doped counterpart, consistent with its insulating nature. Our results point out the dramatic effect of preparation conditions in affecting charge carrier density as well as structural, band and electronic parameters in these systems.

## Introduction

The crystal structure of layered superconductors, such as CuO$_2$-based and Fe-based compounds, is described as consisting of a charge reservoir layer plus a high mobility layer, with the former providing charge doping and the latter hosting the Cooper pairs. This charge transfer scenario has offered a key to understand superconducting mechanisms in these materials and, even more importantly, it has driven the efforts of researchers towards the synthesis of new superconductors. Indeed, a new class of layered superconductors has been discovered very recently, namely BiS$_2$-based compounds such as Bi$_4$O$_4$S$_3$ (T$_c$~4.5 K) [1,2], Sr$_{1-x}$La$_x$FBiS$_2$ [3] and REO$_{1-x}$F$_x$BiS$_2$ (RE=La, Ce, Pr, Nd) systems. Among the latter ones, LaO$_{0.5}$F$_{0.5}$BiS$_2$ exhibits the largest onset T$_c$~11.1 K [4,5]. The crystal structure of La(O,F)BiS$_2$ is composed of a stacking of La$_2$(O,F)$_2$ layers and double BiS$_2$ layers. In terms of electronic bands, the parent compound LaOBiS$_2$ is a band insulator with an energy gap of 0.4 eV between a valence band consisting of O and S *p* states and a conduction band consisting of Bi *6p* and S *3p* states [6]. With the partial substitution of O by F, electrons are doped into the conduction band and eventually superconductivity appears, with maximum T$_c$ of about 3 K for x=0.5, in correspondence of a topological change of the Fermi surface [7,6]. More specifically, the Fermi surface is markedly two-dimensional and close to x=0.5 it fulfills good nesting conditions [6,7]. From calculation of phonon spectrum and electron-phonon coupling constant for LaO$_{0.5}$F$_{0.5}$BiS$_2$ ($\lambda$=0.8) [8], it turns out that this material is a conventional phonon mediated superconductor, however it has been suggested that although the electron-phonon interaction plays the main role in the Cooper pairing, the good Fermi surface nesting may cooperate to give an enhanced attractive pairing interaction around the nesting vectors [7], in analogy with what is generally accepted to occur for iron-based superconductors. This gives rise to an s-wave pairing with a constant gap sign, which has been indeed observed experimentally by muon-spin spectroscopy [9]. Nevertheless, the issues of pairing mechanism and pairing symmetry are still under debate. Indeed, scanning tunneling microscopy studies have found very large values of the reduced gap in BiS$_2$ based superconductors [10], more compatible with unconventional pairing symmetry than with conventional s-wave superconductivity. Moreover, neutron scattering studies have suggested that the electron-phonon coupling in LaO$_{0.5}$F$_{0.5}$BiS$_2$ could be weaker than expected [11]. A more insightful investigation of the relationship between superconducting and normal state properties may add further clues about the phase diagram and thereby also about the conventional versus unconventional character of the relevant pairing mechanism.

In this work, we specifically address the issue of how F doping and preparation conditions, in particular high pressure (2 GPa) annealing, affect normal state properties and superconducting T$_c$ in LaO$_{1-x}$F$_x$BiS$_2$ bulk samples. Indeed, it has been found that high pressure annealing yields a uniaxial lattice contraction along the c axis, as well as elongation of the a axis [12,4], which could be positively linked with the enhancement of superconducting T$_c$ from 3 to 11 K [13,4]. Also the effectiveness of charge doping may be more or less directly affected by the preparation conditions. Theoretical investigations have indicated F doping as the most effective to induce free charge carriers in BiS$_2$ planes, as compared to other possible chemical substitutions [14]. A charge transfer of about 0.26*e* per unit formula to the BiS$_2$ planes in x=0.5 F doped samples has been calculated [6]. On the other hand, F doping has been predicted to affect the band gap [15], making the relationship between nominal and effective doping less straightforward. From the experimental point of view, it has been evidenced that Bi deficiency may severely decrease the actual charge doping with respect to the nominal one [16].

In the following, we present magnetotransport, thermoelectrical and thermal transport characterization of the normal state of LaO$_{1-x}$F$_x$BiS$_2$ bulk samples, in order to gain insight into the electronic properties of this compound. In particular, we examine an undoped sample (x=0) as well as two optimally doped ones (x=0.5) prepared with or without high pressure annealing and compare their behaviors.

**Experimental**

Polycrystalline samples of $LaO_{0.5}F_{0.5}BiS_2$ are prepared by solid-state reaction using powders of $La_2S_3$ (99.9%), $Bi_2O_3$ (99.9%), $BiF_3$(99.9%), $Bi_2S_3$ and grains of Bi (99.99%). The $Bi_2S_3$ powder is prepared by reacting Bi (99.99%) and S (99.99%) grains at 500 ºC in an evacuated quartz tube. The starting materials with a nominal composition of $LaO_{0.5}F_{0.5}BiS_2$ are mixed-well, pressed into pellets, sealed into an evacuated quartz tube and heated at 700 ºC for 10h. As an additional step, the obtained polycrystalline samples are annealed at 600 ºC under a high pressure of 2 GPa for 1 hour, using a cubic-anvil high-pressure synthesis apparatus.

Magnetotransport behavior is investigated by means of a Physical Properties Measurement System (PPMS) by Quantum Design at temperatures from room temperatures down to 2K and in magnetic fields up to 9T. Hall coefficients ($R_H$) are determined measuring the transverse resistivity at selected fixed temperatures, sweeping the field from -9T to 9T. Seebeck (S) effect and thermal conductivity are measured with the PPMS Thermal Transport Option in continuous scanning mode with a 0.3 K/min cooling rate.

**Results**

In figure 1, the resistivity ρ curves of the undoped and doped samples prepared either by conventional solid-state reaction at ambient pressure ($LaO_{0.5}F_{0.5}BiS_2$-LP) or by high-pressure annealing ($LaO_{0.5}F_{0.5}BiS_2$-HP) are presented. All the three samples exhibit semiconducting behavior in the whole temperature range of the normal state. The values of the room temperature resistivity are 28 mΩcm for the undoped sample and an order of magnitude smaller for the doped ones, namely 3.2 mΩcm for $LaO_{0.5}F_{0.5}BiS_2$-LP and 4.8 mΩcm for $LaO_{0.5}F_{0.5}BiS_2$-HP, all in fair agreement with previous reports [5,12,13,17]. Despite the two doped samples $LaO_{0.5}F_{0.5}BiS_2$-LP and $LaO_{0.5}F_{0.5}BiS_2$-HP have similar resistivity values at low temperature, the $LaO_{0.5}F_{0.5}BiS_2$-LP sample curve has a steeper temperature dependence, suggesting larger activation energy for transport. The doped sample $LaO_{0.5}F_{0.5}BiS_2$-LP become superconducting at temperature $T_{c-90\%}$=2.36K (defined at 90% of normal state resistivity) and the resistivity is not completely vanished at the minimum temperature of our measurement T=2K. In the case of the doped sample $LaO_{0.5}F_{0.5}BiS_2$-HP we find a transition temperature $T_{c-90\%}$=8.46K and a transition width of 1.24K. These data are again in agreement with previous works reporting on the effect of high pressure annealing in improving superconducting properties [4,13].

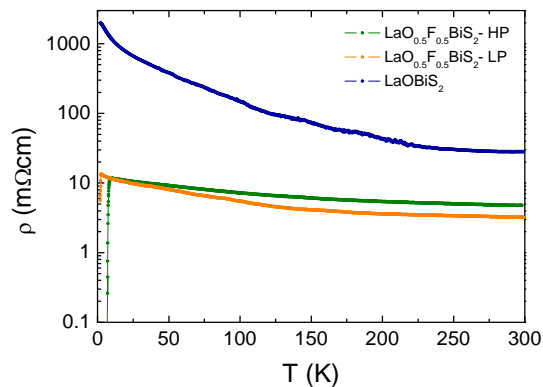

**Figure 1: (color online)** resistivity versus temperature curves of undoped $LaOBiS_2$ and doped $LaO_{0.5}F_{0.5}BiS_2$ polycrystals, prepared either with or without high pressure annealing.

The normal state longitudinal magnetoresistivity (ρ(H)-ρ(H=0))/ρ(H=0), not shown, is negligibly small (<1%) at all temperatures for all the samples.

In figure 2, Hall effect curves of the three samples are plotted. In the case of the doped $LaO_{0.5}F_{0.5}BiS_2$-HP sample, $R_H$ is negative at all temperatures and slightly increasing in magnitude with decreasing temperature, with a value around $-10^{-8}$ m$^3$/C at low temperature just above the superconducting transition. This behavior is indeed reminiscent of that of Fe-based superconductors of different families, e.g. F doped 1111 compounds [18,19], electron doped 122 compounds [20,21] and other families of Fe-based superconductors [22]. For the $LaO_{0.5}F_{0.5}BiS_2$-LP sample the magnitude of $R_H$ is on average 5 times as much, consistent with other literature data [23], indicating lower electron density. However, quite remarkably, it is positive in sign. Positive $R_H$ has also been found in other $BiS_2$ superconductors at optimal electron doping [24]. Finally, $R_H$ of the $LaOBiS_2$ sample is negative, weakly dependent of temperature and its magnitude is around few times $10^{-7}$ m$^3$/C, i.e. 10 times larger than that of the corresponding doped $LaO_{0.5}F_{0.5}BiS_2$-LP sample.

It must be said that whereas for the $LaO_{0.5}F_{0.5}BiS_2$-HP sample Hall effect measurements are carried out easily, in the case of the $LaO_{0.5}F_{0.5}BiS_2$-LP and $LaOBiS_2$ samples the experimental data points are more scattered, not due to the magnitude of the Hall signal itself, but possibly due to charging at grain boundaries. Also bismuth and bismuth oxide impurities may be responsible of erratic Hall effect measurements, but in our case bismuth phase is not detected by X-rays diffraction analysis and moreover the negligible magnetoresistance is not consistent with magnetotransport behavior of bismuth and bismuth oxides.

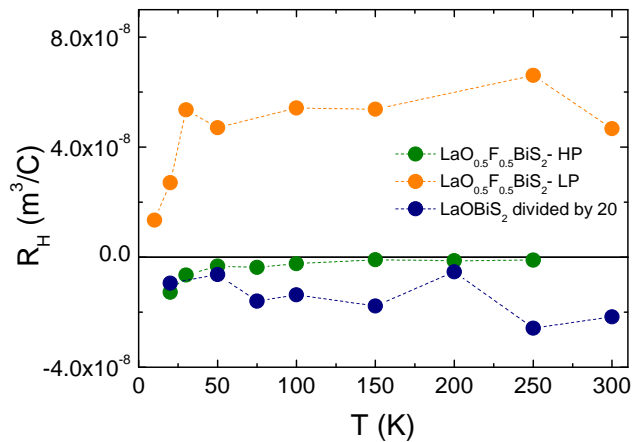

**Figure 2: (color online)** Hall coefficient versus temperature curves of undoped $LaOBiS_2$ and doped $LaO_{0.5}F_{0.5}BiS_2$ polycrystals, prepared either with or without high pressure annealing.

In figure 3, we present Seebeck coefficient S curves of the three samples. It can be seen that in all cases, the Seebeck coefficients are roughly linear with temperature, suggesting that the diffusive mechanism of charge carriers dominates. This result is in agreement with similar optimally doped compounds such as $PrO_{0.3}F_{0.7}BiS_2$ [25], but at odds with other $BiS_2$-based superconductors, where non-monotonic behavior with a pronounced minimum possibly related to a drag mechanism is observed [26]. Contrarily to Hall resistance curves, the Seebeck coefficient curves are negative at all temperatures for all the three samples. In particular, the $LaO_{0.5}F_{0.5}BiS_2$-LP exhibits negative S and positive $R_H$, similarly to the case of other electron-doped $BiS_2$ superconductors [24]. No appreciable dependence on magnetic field is detected in the normal state Seebeck curves (in-field curves are not shown in figure 3), as expected in the diffusive regime. Only below $T_c$ of the doped samples the Seebeck curves at $\mu_0H=7T$ and 0T deviate, as a consequence of the finite upper critical field.

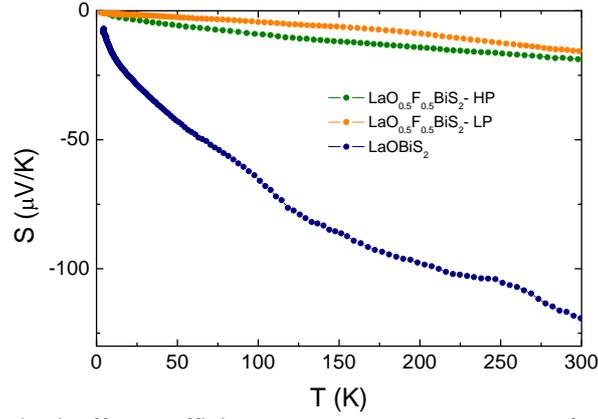

**Figure 3: (color online)** Seebeck effect coefficient versus temperature curves of undoped LaOBiS$_2$ and doped LaO$_{0.5}$F$_{0.5}$BiS$_2$ polycrystals, prepared either with or without high pressure annealing.

In figure 4, thermal conductivity κ curves of the three samples are displayed. For the undoped sample, the thermal conductivity increases with temperature at low temperature, exhibits a maximum around T=100K and levels off at higher temperatures. On the other hand, both the doped samples exhibit monotonic thermal conductivity and smaller thermal conductivity values. The doped LaO$_{0.5}$F$_{0.5}$BiS$_2$-HP sample has smaller thermal conductivity than the LaO$_{0.5}$F$_{0.5}$BiS$_2$-LP sample. No dependence of κ on magnetic field is observed (in-field curves are not shown in figure 4).

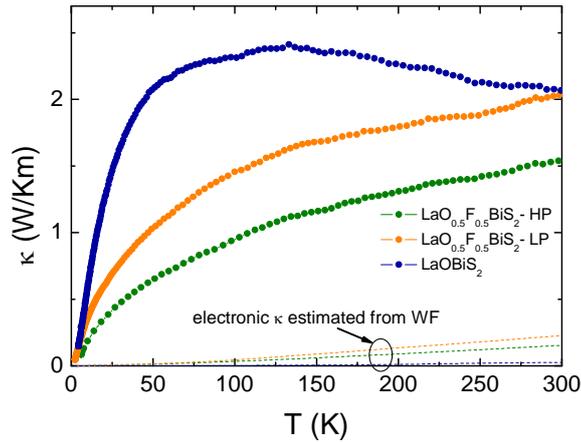

**Figure 4: (color online)** thermal conductivity versus temperature curves of undoped LaOBiS$_2$ and doped LaO$_{0.5}$F$_{0.5}$BiS$_2$ polycrystals, prepared either with or without high pressure annealing. Dashed lines indicate the electronic contribution to the thermal conductivity estimated with the Wiedemann-Franz (WF) law.

**Data analysis and discussion**

Combining resistivity, Hall coefficient and Seebeck coefficient data we can carry out a fit of normal state properties, with carrier densities $n$, mobilities $\mu$ and effective masses $m_{eff}$ as parameters. In a single band picture we use for resistivity and Hall coefficient the relationships $\rho=(en\mu)^{-1}$ and $R_H=(en)^{-1}$, while for the Seebeck coefficient we use the Mott relationship valid for isotropic degenerate materials in the diffusive regime:

$$S = -\frac{\pi^2}{3}\left(\frac{k}{e}\right)kT\left(\frac{d\ln\sigma(E)}{dE}\right)\bigg|_{E_F} \approx -\left(\frac{3}{2}+\alpha\right)\frac{8\pi^{8/3}k^2}{3^{5/3}h^2 e}m_{eff}\frac{T}{n} \qquad (1)$$

where k is the Boltzmann constant, $h$ the Planck constant, $E$ is the energy of charge carriers, α characterizes the energy dependence of the scattering time $\tau\sim E^\alpha$ (which is assumed constant for simplicity, i.e. $\alpha\sim 0$), σ is the conductivity ($\sigma=\rho^{-1}$) and the logarithmic derivative is calculated at the Fermi level. The minus sign holds for n type charge carriers. To get the above expression in terms of fitting parameters $n$ and $m_{eff}$, a free electron picture with parabolic dispersion and spherical Fermi surface is assumed. In case of two band picture, with electrons (subscript e) and holes (subscript h) contributing to transport, the appropriate relationships would be $\rho=((en_e\mu_e)^{-1}+(en_h\mu_h)^{-1})^{-1}$, $R_H = \frac{1}{e}\frac{(\mu_h^2 n_h - \mu_e^2 n_e)}{(\mu_h n_h - \mu_e n_e)^2}$ and $S = \frac{\sigma_h S_h + \sigma_e S_e}{\sigma_h + \sigma_e}$, where the Seebeck coefficients of each band are calculated according to the Mott relationship eq. (1).

Clearly in a *single band approximation* these parameters are univocally determined, whereas if we assume both electrons and holes contributions the solution are not anymore univocal. In figure 5 we show the mobility and carrier concentration resulting from the single band fitting of the LaO$_{0.5}$F$_{0.5}$BiS$_2$-HP sample. The electron mobility is on average ~0.5-2cm$^2$/(Vs), slightly decreasing with increasing temperature. The electron density is in the range $5\cdot10^{20}$-$7\cdot10^{21}$ cm$^{-3}$, slightly increasing with increasing temperature. Noteworthy, $2.3\cdot10^{21}$ cm$^{-3}$ is just what expected from the nominal x=0.5 F doping. Moreover, in ref. [6] it is predicted that 0.7+0.26=0.96 electrons per unit cell are doped in the BiS$_2$ layers of LaO$_{0.5}$F$_{0.5}$BiS$_2$, corresponding to $4.4\cdot10^{21}$ cm$^{-3}$ in fair agreement with our data. However these charge carriers seem not to be very mobile, as seen from low mobility values and from semiconducting behavior of resistivity, possibly as a consequence of strong interaction with boson fluctuations, or else as a consequence of many-body interactions [7], or else in relation with the predicted quasi-nested Fermi surface leading to the charge density wave instability [27]. From the fitting, values of the electron effective mass around $m_0$ ($m_0$ bare electron mass) are obtained, in agreement with ref. [26].

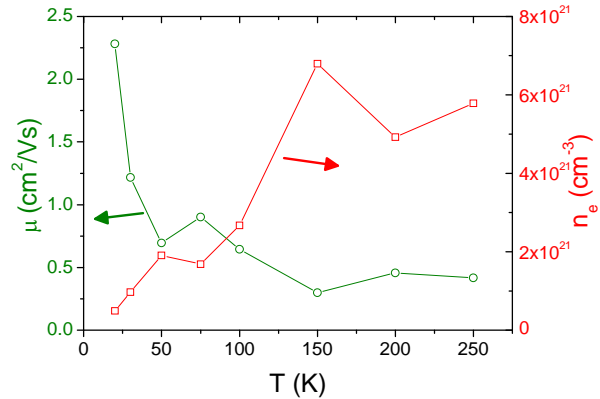

**Figure 5: (color online)** electron charge density and mobility obtained from single band fitting of $\rho$ and $R_H$ data of doped LaO$_{0.5}$F$_{0.5}$BiS$_2$_HP sample, prepared with high pressure annealing.

If we assume a *two band description* with hole and electron contributions, all the physically realistic solutions for the set of parameters present some common features: (i) electron mobilities are larger than hole mobilities at all temperatures; (ii) at low temperature electron densities are around $10^{21}$ cm$^{-3}$ or larger and holes densities are one order of magnitude smaller than electron density, while at high temperature hole and electron densities are similar; (iii) effective masses are of the order of $m_0$. This picture indicates that below 100K the system can be approximately described in a single band picture, taking into account only the electron contribution to transport.

For the doped LaO$_{0.5}$F$_{0.5}$BiS$_2$_LP sample, it is not possible to assume a single band description, because it could not account for the different sign of $R_H$ and S. On the contrary, in a two band description this situation may occur, because, roughly speaking, bands contributions are weighed by $n\mu^2$ in the expression for $R_H$, while they are weighed by $\mu$ in the expression for S. Hence, if we allow for a band of electrons with smaller density and higher mobility, it may determine the negative sign for S even with a positive $R_H$. All the different solutions compatible with our experimental data present common robust characteristics, namely: (i) both electron and hole mobilities increase with increasing temperature; $\mu_e$ varies from few cm$^2$/(Vs) at low temperature to several tens cm$^2$/(Vs) at 300K, while $\mu_h$ is one order of magnitude smaller; (ii) both electron and hole densities are weakly temperature dependent except for the lowest temperature; $n_e$ is $1\text{-}5\cdot10^{18}$ cm$^{-3}$, while $n_h$ is $10^{19}\text{-}10^{20}$ cm$^{-3}$; (iii) both effective masses are around 0.5-1 $m_0$. In short, according to this picture there are more holes than electrons, but holes have much lower mobility. It may appear strange that the LaO$_{0.5}$F$_{0.5}$BiS$_2$_LP sample has lower |S| and at the same time lower carrier concentrations as compared to the LaO$_{0.5}$F$_{0.5}$BiS$_2$_HP sample. Indeed, this is a multiband effect, explained by the more balanced competition between electron and hole type carriers in the former sample.

Finally single band fittings for the undoped LaOBiS$_2$ sample yield average electron density few times $10^{19}$ cm$^{-3}$, a factor 10 smaller than that of the corresponding doped sample LaO$_{0.5}$F$_{0.5}$BiS$_2$_LP, mobilities values that increase with temperature from 0.2 to 15 cm$^2$V$^{-1}$s$^{-1}$ and effective masses around ~0.5$m_0$. Obviously a two band scenario is possible as well.

In summary, the simple inspection of rough experimental data of resistivity, Seebeck coefficient and $R_H$ in LaO$_{0.5}$F$_{0.5}$BiS$_2$_LP and LaO$_{0.5}$F$_{0.5}$BiS$_2$_HP samples, reveal that these two samples are significantly different. Indeed, $R_H$ of the LaO$_{0.5}$F$_{0.5}$BiS$_2$_LP sample is positive and nearly 5 times larger in magnitude than $R_H$ of the LaO$_{0.5}$F$_{0.5}$BiS$_2$_HP sample, which is negative. S of the two samples are both negative, but the S magnitude in the LaO$_{0.5}$F$_{0.5}$BiS$_2$_LP sample is slightly smaller than S of the LaO$_{0.5}$F$_{0.5}$BiS$_2$_HP sample. Finally, resistivities are comparable in magnitude, but the curve of the LaO$_{0.5}$F$_{0.5}$BiS$_2$_LP sample has steeper temperature dependence, indicating larger activation energy in the LaO$_{0.5}$F$_{0.5}$BiS$_2$_LP sample. As a further step, the data analysis, despite its simplifying assumptions, confirms that the preparation conditions have a strong effect on normal state properties. The average values of the parameters obtained from the fittings in the three samples are summarized in Table I, for easy comparison. The major difference between LaO$_{0.5}$F$_{0.5}$BiS$_2$_LP and LaO$_{0.5}$F$_{0.5}$BiS$_2$_HP samples is the enhanced electron density in the latter, which is likely closely related to the enhancement of superconducting $T_c$ from 3 to 11 K [4,13]. Moreover, these data suggest that the preparation conditions yield not only different charge carrier densities, but also different band and electronic parameters, such as larger effective mass and lower mobility in the LaO$_{0.5}$F$_{0.5}$BiS$_2$_HP sample as compared to the LaO$_{0.5}$F$_{0.5}$BiS$_2$_LP sample. Both the larger effective mass and the lower mobility may be possibly related to stronger coupling with boson excitations, indeed enhanced coupling strength at optimal doping has been suggested for BiS$_2$ based superconductors in ref. [24]. We note that also the analysis of superconducting properties has indicated different band and electronic parameters in samples prepared either with or without high pressure annealing [28].

|  | LaO$_{0.5}$F$_{0.5}$BiS$_2$_HP (single band fit) | LaO$_{0.5}$F$_{0.5}$BiS$_2$_LP (two band fit) | LaOBiS$_2$ (single band fit) |
|---|---|---|---|
| **charge density (cm$^{-3}$)** | $5\cdot10^{20}\text{-}7\cdot10^{21}$ | electrons: $1\text{-}5\cdot10^{18}$<br>holes: $7\cdot10^{19}\text{-}1.6\cdot10^{20}$ | $1\text{-}5\cdot10^{19}$ |
| **charge mobility (cm$^2$V$^{-1}$s$^{-1}$)** | 0.5-1 | electrons: 10-100<br>holes: 3-25 | 0.2-15 |
| **effective mass** | ~1$m_0$ | 0.5-1$m_0$ | ~0.5$m_0$ |

**Table I:** parameters obtained by fitting of $\rho$, $R_H$ and S curves

We now consider the thermal conductivities of the three samples. For the undoped sample, the thermal conductivity increases with temperature at low temperature, due to increasing excitation of phonons. In this regime the dominant phonon scattering mechanism is by defects and grain boundaries. At high temperature, the thermal conductivity decreases with temperature, due to enhanced phonon-phonon scattering, which is the dominant scattering mechanism in this regime. At intermediate temperature there is a broad maximum, as a consequence of the crossover between the above mentioned regimes. It is expected that with increasing disorder or grain boundary density, the regime of defect scattering extends up to larger temperatures, and eventually the broad maximum is damped, yielding a monotonic shape of the κ(T) curve. This is indeed the case of the F doped samples, which show monotonic behavior typical of more disordered materials (or materials with higher density of grain boundaries) as compared to the undoped sample, which seems to have a more ordered crystal lattice (or larger grains).

It may appear unexpected that the doped sample prepared at high pressure has smaller thermal conductivity than the one prepared at low pressure, given that the former should have higher density and better intergrain connection. On the other hand, this finding is consistent X-rays analysis [4], which exhibits more broadened X-rays peaks after high pressure annealing. The broader peaks in high pressure annealed samples indicate the presence of disorder and inhomogeneity, which may be due to a competition of tetragonal and monoclinic phases in high pressure samples [29]. Structural inhomogeneity in similar samples has been also detected from the analysis of magnetotransport data in the superconducting state, where the upper critical field parallel to BiS planes has been found to be anisotropic, indicating that these crystalline planes are distorted into a lower-symmetry as compared to the tetragonal structure [28]. It can be noted that also electronic mobility, which is smaller in the $LaO_{0.5}F_{0.5}BiS_2\_HP$ sample (see Table I), may be affected by such disorder and inhomogeneity.

Using the Wiedemann-Franz law we estimate that the thermal conductivity is dominated by phonon thermal transport for all the samples, as indicated by the dashed lines in figure 4. This finding is consistent with the low charge carrier mobility (in all samples κ is larger than 1.5 W/Km at room temperature, while the corresponding electron contribution is estimated smaller than 0.2 W/Km). The absence of magnetic field dependence of κ is consistent with κ being dominated by the lattice contribution.

**Conclusions**

We study normal state electrical, thermoelectrical and thermal transport in polycrystalline undoped $LaOBiS_2$ and doped $LaO_{0.5}F_{0.5}BiS_2$ samples, prepared either by conventional solid-state reaction at ambient pressure or by additional high pressure annealing.

All our samples present common behaviors such as semiconducting resisitivity, diffusive and negative Seebeck effect and thermal transport dominated by phonon contribution. However, the major difference between doped $LaO_{0.5}F_{0.5}BiS_2$ samples, prepared either by conventional solid-state reaction at ambient pressure or by high pressure annealing, lies in the sign and magnitude of the Hall resistance, which is positive in sign and larger in magnitude in the former sample and negative in sign and smaller in magnitude in the latter sample. This finding indicates the role played by the multiband character of these compounds. The higher electron concentration of samples prepared by high pressure annealing may be closely related to the enhanced superconducting $T_c \sim 11$ K and to the change of structural parameters. Our results indicate that not only charge carrier densities, but also band and electronic parameters are affected by preparation conditions. This work should trigger *ab initio* calculations which could provide a theoretical framework for our experimental outcomes.

**Acknowledgments**

The work was supported by the FP7 European project SUPER-IRON (contract No. 283204).